# Experimental Investigation of an Incremental Contact Model for Hyperelastic Solids Using In-Situ Optical Interferometric Technique

Chunyun Jiang[2], Yanbin Zheng[1, *]

[1] School of Science, Lanzhou University of Technology, Lanzhou 730050, China

[2] Department of Engineering Mechanics, SVL and MMML, Xi'an Jiaotong University, Xi'an 710049, China

* Correspondence: zhengyb@lut.edu.cn

**Abstract**: The hyperelastic materials would contribute to the intricacies of rough surface contact, primarily due to the heightened nonlinearity caused by stress concentration. In our previous research, an incremental contact model tailored for hyperelastic materials is proposed and validated by finite element (FEM) simulations. From an experimental perspective, this study employs an in-situ optical interferometric technique to precisely document the actual contact zone between hyperelastic solids and quartz glass. Simultaneously, the contact force is meticulously recorded in sync by a force sensor positioned beneath the hyperelastic samples. Comparing with the predictions of incremental contact model for hyperelastic materials, a significant agreement becomes evident, almost in a range of nearly complete contact. Its significance extends to practical domains such as sealing mechanisms, leakage prevention, and structural integrity, offering valuable insights for these applications.

---

[1] Corresponding author
E-mail: zhengyb@lut.edu.cn

# 1 Introduction

The contact of hyperelastic solids like rubber is ubiquitous in everyday life and industrial production, encompassing applications such as tires, seals, cables, and more [1]. It is worth noting that the nonlinear behavior of hyperelastic materials would be magnified by stress concentration due to microscale contacts and render the study of hyperelastic solid contact behavior significantly more complex. So far, investigating the impact of hyperelasticity on contact behavior represents a crucial challenge within academic fields [2-5].

Incorporating surface roughness into contact models presents a formidable challenge due to the intricate nature of randomness and features over multiple scales. Currently, two primary characterizations of rough surface morphology prevail: statistical description and fractal description. In the early stages of research, Greenwood and Williamson [6] established the classical statistical contact model in 1996, known as the GW model. This model was groundbreaking in establishing a linear relationship between contact load and actual contact area by employing Hertz's solution. Based on the GW model, various statistical models have since emerged, each incorporating different hypotheses. These hypotheses include considerations for nonuniform asperity radii [7-9], elliptic paraboloidal asperities [10], and anisotropic topographies [11]. As to fractal description, in 1982, Mandelbrot [12] found, most rough surfaces in the nature can be described as self-affine fractals. This self-affinity property signifies that these surfaces maintain statistical equivalence when their height ($h$) and lateral coordinates ($x$ and $y$) are rescaled by varying factors. Building upon fractal theory, Persson [13] created a theory for rough surface contact utilizing power spectral density (PSD) and assuming the elasticity of rubber materials. Recently, Wang [14] adopted a deterministic description of rough surface and put forward an incremental contact model for rough surfaces. This model was validated by the finite element method[14, 15] (FEM) and experiments [16], focusing on contact area fractions within the 15% range. These aforementioned contact theories could contribute to the realization in mechanism of rough surfaces.

Besides morphology descriptions and contact mechanism, the property of materials is one of crucial factors affecting the contact response [17]. In many studies, FEM simulations offer an alternative and convenient method to investigate rough surface contact considering complex influencing factors, for example the nonlinearity of materials. Song et al. [18], Zhang et al. [19] and Jiang et al. [20] studied the contact behaviors of rough surfaces by taking the size dependence into account. Zhang and Yang [21] noted that the indentation behaviors of hyperelastic spheres primarily depend on the combined influences of substantial deformation and material nonlinearity. By introducing the instantaneous tangent modulus $E_t$, Jiang et al. [22] extended the incremental model into the hyperelasytic materials and this extension was subsequently validated through FEM. Similarly, Lengiewicz et al. [23] emphasized that hyperelasticity introduces notable differences in the contact deformation process under high loads, as observed through FEM analysis. All these researches indicate the nonlinearity of materials plays as a significant role in deformation. Therefore, considering hyperelasticity in the context of rough surface contact seems to be a reasonable approach.

In addition to theoretical investigations and numerical simulations, the evolution of interfacial contact can also be gleaned through experimental observations. The physical technologies, such as thermal resistance [24], electric resistance [25], and X-ray examination [26], and ultrasound reflection [27-29] at the contact interface, can exhibit noteworthy changes in response to variations in actual contact area at the interfaces. In addition to abovementioned methods, the optical technique [30-34] have also been utilized to investigate the rough surface contact with the advantage of in-site measurement and direct observation. Liang et al [16] and Li et al. [34] utilized the frustrated total internal reflection technique (FTIR) for studying the elastoplastic deformation of metals which possess super catoptric performance. As an alternative optical technique, Hertz pioneered the use of interferometric techniques to measure surface separation and established the foundation for the field of contact mechanics [35]. For polymer materials, Krick et al. [31] employed the 0th order interference fringe to identify the actual contact regions, and Benz et al. [36] utilized optical interferometric analysis to measure polymer deformation at the contact interface. Compared to FTIR, the application of optical interferometric technique results in reduced light pollution and enhances the detection of small contact regions.

In this paper, we have conducted uniaxial tensile (UT), planar tensile (PT), and biaxial tensile (BT) experiments to comprehensively determine the mechanical property of the hyperelastic material. And, the 6th-order-Ogden constitutive model was selected and fitted in commercial software ABAQUS with the experimental data. Subsequently, rough surface contact experiments were conducted using an interferometric method and compared with the incremental model for hyperelastic materials [22] and both results reach great agreement within a contact fraction range of 90%. Furthermore, it is found that the incremental model demonstrated a strong predictive capability for the contact behavior of hyperelastic rough surface. At the same time, the application of tangent modulus introduced the nonlinearity of material. The ratio of the tangent modulus to the linear elastic modulus varies with contact stresses and consistently falls within the range of approximately 2.2 ~ 3, which aligns with the FEM results presented in the reference [22]. This research experimentally demonstrates that the influence of the material's nonlinearity on contact behavior, driven by stress concentration at the contact surface, enhances the instantaneous tangent modulus of the material at the contact interfaces.

## 2 The experiments of material constitutive model

Selecting a proper material constitutive model and precise parameterization are pivotal factors in accurately capturing mechanical deformation. Because the strain energy density (SED) functions for hyperelastic materials are formulated based on phenomenological or statistical theories, their functional expressions exhibit variability and involve distinct parameters. In order to exactly represent the mechanical properties of the materials in the deformation, UT, PT, and BT experiments are required [5]. The rubber samples for UT, PT, and BT experiments are crafted from the same sheet of ethylene-propylene terpolymer (EPDM). This material is generally applied in waterproof materials, cable sheaths, heat-resistant rubber pipes and sealings. The UT experiment employed a standard dumbbell-shaped specimen measuring 6mm × 115mm × 1.5mm. For the PT experiment, a rectangular specimen with dimensions of 40mm × 10mm × 1.5mm was employed, while the BT experiment utilized a cross-shaped specimen measuring 40mm × 12mm × 1.5mm. The UT equipment is WANCE@ETM104B, featuring a maximum sensor range of 10kN and a minimum resolution of 0.01N. The PT equipment, identified as EUM-25k25, utilizes a sensor with a maximum range of 3kN and a minimum resolution of 0.1N. The biaxial tensile equipment is designated as IPBF-300, featuring a sensor with a maximum range of 300N and a minimum resolution of 0.01N.

Three different types of experimental samples and the tensile testing equipment is illustrated in Fig. 1. The tensile load could be continuously monitored in real-time via the force sensor, whereas strain measurements required calculations involving the continuous tracking of markers on the samples through digital image processing technology, a technique known as non-contact strain measurement technology. This method is commonly employed in digital image correlation (DIC) for the purpose of capturing related information. Sprayed speckle markers are applied to the samples' surface to instantaneously capture the movement and deformation of speckles using an in-situ camera. This method offers several advantages, including minimal demands on the experimental environment, non-contact whole-field measurement capabilities, robust resistance to interference, and high measurement precision. Consequently, we employ the software MATLAB to perform images processing and calculation of the elongation ratio and strain.

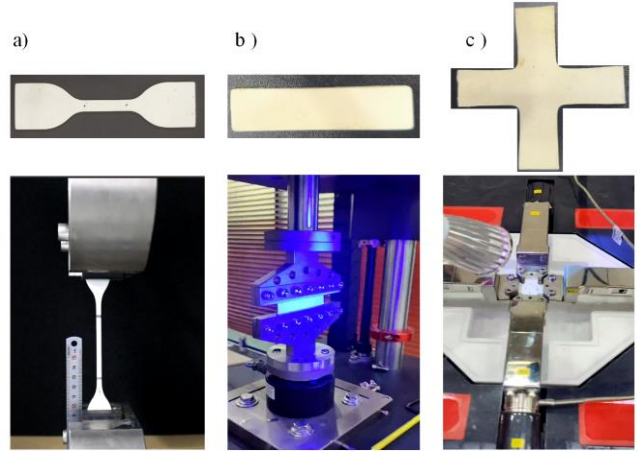

Fig. 1 Rubber specimens and tension setup for (a) UT, (b) PT, and (c) BT experiments

As depicted in Fig. 2, the symbols represent the stress-strain data obtained through experiments, and the dashed lines represent the fitting data. Using one proper SED function to satisfy the three experimental data simultaneously is a challenge on the form of the function. To better characterize the mechanical properties of the materials, the three kinds of tensile experimental data are fitted simultaneously in the commercial software ABAQUS. By employing various SED functions for fitting and conducting comparative analyses, we determined that the 6th-order-Ogden SED function is the most appropriate choice for the experimental data. The SED function is written as

$$W = \sum_{i=1}^{6} \frac{2\mu_i}{\alpha_i^2}(\bar{\lambda}_1^{\alpha_i} + \bar{\lambda}_2^{\alpha_i} + \bar{\lambda}_3^{\alpha_i} - 3), \tag{1}$$

where $\mu_i$ and $\alpha_i$ are material constants, listed in Table 1. The stretch $\lambda_t = 1+\varepsilon_t$ is a function of strain $\varepsilon_t$ And the initial shear modulus $\mu_0 = \sum_{i=1}^{6} \mu_i$. To facilitate the normalization, here, we give the initial Young's modulus $E_0 = 2\mu_0(1+\nu)$ and the initial composite modulus $E_0^* = E_0/(1-\nu^2)$. For incompressible materials, the Poisson ratio $\nu = 0.5$ in this paper. The stress-strain curve could be derived by invoking the principle of virtual work [37], donated as

$$S = \phi(\varepsilon) = \sum_{i=1}^{6} \frac{2\mu_i}{\alpha_i}[(1+\varepsilon)^{\alpha_i - 1} - (1+\varepsilon)^{-2\alpha_i - 1}], \qquad (2)$$

where, $S$ represents the Piola-Kirchhoff stress.

Table 1 The parameters of the 6th-order-Ogden SDE function

| Parameters | 1 | 2 | 3 | 4 | 5 | 6 |
|---|---|---|---|---|---|---|
| $\mu_i$ /Mpa | 5025.77 | -3862.35 | 1442.35 | -4644.15 | 3119.98 | -1079.81 |
| $\alpha_i$ | 0.635 | 0.976 | 1.178 | 0.259 | -0.0488 | -0.219 |

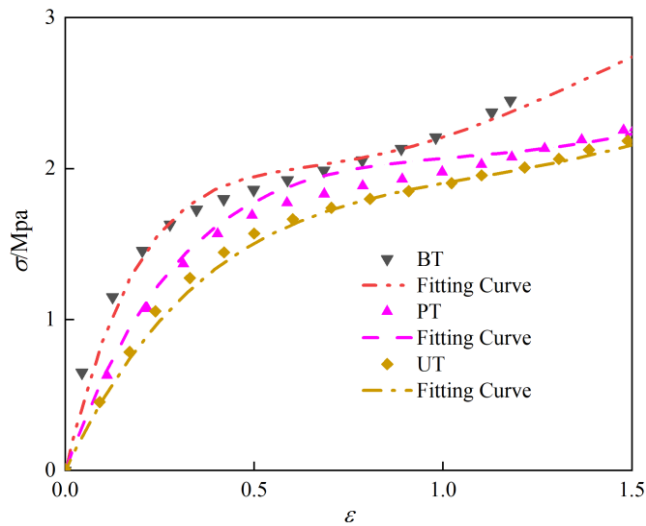

Fig. 2 The experimental data (lines) correspond to UT, PT, and BT experiments, and the fitting data (symbols) obtained using Eq. (2)

## 3 The contact experiments of hyperelastic rough surface

The optical interferometric technique [31, 36] has been adopted to measure the actual contact areas of hyperelastic materials. In this paper, we built an analogous experimental setup to conduct rough surface contact experiments using optical interferometric technique. Fig. 3 illustrates the schematic of the experimental setup for rough surface contact, comprising primarily three integral components: the loading system, the contact module, and the camera capture system.

In the loading system, the sample of rough surface is positioned on a force sensor (FA703, SIMBATOUCH) with data acquisition rate of 50 Hz and a resolution of 0.01 N. The force sensor will record the variation of contact force. These components are driven by a servo motor with a minimum loading rate of 3μm/s, which is slow enough to guarantee quasi-static state and reduce the influence of material viscoelastic relaxation on the contact area. A high-quality scientific complementary metal oxide semiconductor (CMOS) camera (2048 × 2048 pixels, with a pixel size of 6.5 μm × 6.5 μm and an impressive 16-bit gray depth, specifically the pco.panda 4.2C model) is positioned vertically above the rough surface samples for precise image capture. To ensure optimal image quality, the camera's exposure time is finely tuned to 50 ms. To generate thin-film interference, A white coaxial illuminant (TZ-D5W) delivers a beam of light. Then, the light could be reflected to passe through objective utilizing an internal mirror in CMOS lens, and illuminates vertically onto the lower surface of the quartz glass and the surface of the sample. The light reflected at both interfaces will undergo interference. Due to the presence of half-wave losses [31], the image captured by CMOS will exhibit a notably darker appearance in the actual contact regions. Subsequently, the actual contact area and the numbers of contact patches could be determined based on the intensity of pixels by employing the digital image processing in MATLAB [16].

In experiments, we control the movement of sample to achieve a complete contact with quartz glass according to abovementioned experimental process. After synchronizing the data, the relationship between the actual contact area $A$ and the contact load $F$ is obtained.

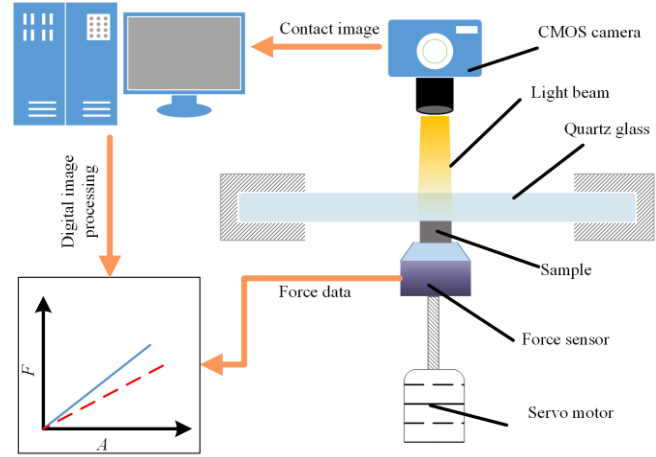

Fig. 3 The schematic of experimental setup for rough surface contact

The morphologies of the rough surface were meticulously scanned using the white light interferometer (NanoMap-1000WLI, AEP) with a vertical resolution of 0.01 nm before contact experiments. Installed with a 10X interference objective lens, each scanning procedure captures a projection area measuring 1047.9 μm × 1047.9 μm. Through an automatic stitching process and filtered to have a lateral resolution of 20μm, the complete surface morphology can be reconstructed from multiple individual scans, as shown in Fig. 4. Through fast Fourier transformation, the PSDs $C(q)$ in $x$ and $y$ directions of four rough surfaces are displayed in the subplots e, f, g, h, i, j, k, l, respectively. It becomes evident that all four surfaces exhibit self-affine fractal characteristics. For each sample, the lowest surface node is selected as the reference height of the $z$-coordinate. By employing a virtual plane to truncate the morphologies and disregarding the deformation of non-contact regions, we can determine the actual contact area faction $A/A_0$ varying with the distance $z/\sigma$ between the virtual plane and the reference plane, as illustrated in the subplots m, n, o, and p. The $A_0$ and $\sigma$ represent the nominal area and roughness of surface, respectively

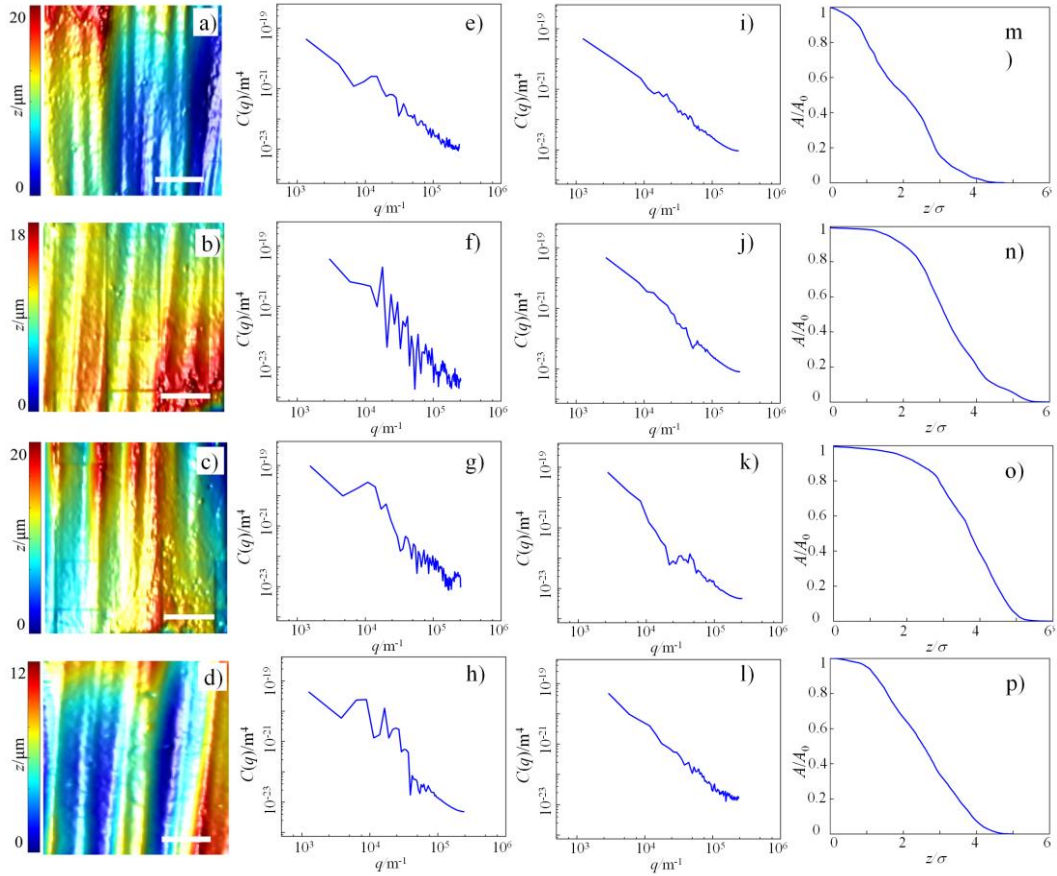

Fig. 4 (a, b, c, d) The morphologies, the PSD in (e, f, g, h) *x* direction and in (i, j, k, l) *y* direction and (m, n, o, p) the evolution of actual contact area by truncation of four samples T1, T2, T3 and T4, respectively. The scale bar corresponds to 1mm.

**4 The incremental equivalent circular contact model for hyperelastic materials**

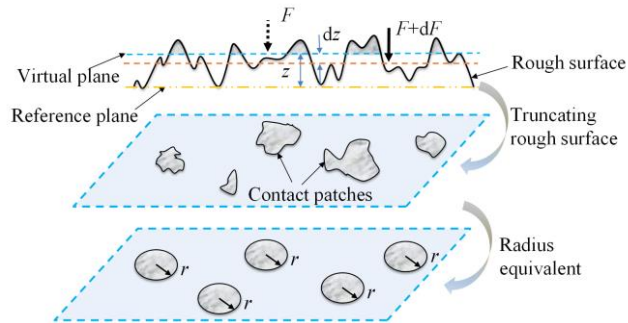

Fig. 5 The schematic of incremental contact model based on profile theory

Fig. 5 illustrates the schematic of the incremental equivalent circular contact model. Here, the lowest node of rough surface is still selected as the reference height of the z-coordinate. Following the profile theory outlined by Abbott and Firestone [38], using a virtual plane to truncate the morphologies of a rough surface, we can extract the actual contact area $A(z)$ and the number of contact patches $N(z)$. $z$ represents the separation between virtual plane and reference plane. While it's important to note that the profile theory doesn't account for finite deformation elastic coupling effects, it remains a valuable tool for estimating the actual contact area[14, 15, 39]. For convenience, an averaging assumption is used to calculate the radius $r(z)$ of contact patches with maintaining $A(z)$ and $N(z)$ constant at separation $z$. Then the radius $r(z)$ could be writen as

$$r(z) = \left[\frac{A(z)}{\pi N(z)}\right]^{1/2}. \tag{3}$$

For the contact of a circular flat rigid indenter on infinite substrate, the contact stiffness has been derived by Sneddon[40], denoted as $2E^*r$. $E^* = E/(1-\nu^2)$ represents the composite elastic modulus, with $E$ representing the Young's modulus of the substrate. Consequently, by multiplying $N(z)$, the stiffness of contact interface at separation $z$ can be expressed as following

$$\frac{dF(z)}{dz} = 2E^* r(z) N(z). \tag{4}$$

In terms of hyperelastic materials, the stress-strain curve exhibits high nonlinearity, especially in compression. The nonlinearity would cause the tangent modulus to fluctuate with the stress, rendering the use of the initial elastic modulus for calculating contact stiffness inappropriate. Analogous to buckling theory [41], the tangent modulus $E_t$ is suggested in the compressive research for hyperelastic materials. Therefore, we introduce the composite tangent modulus $E_t^* = E_t/(1-\nu^2)$ into the incremental contact model to replace $E^*$. Then, Eq. (4) can be rewritten as

$$\frac{dF(z)}{dz} = 2E_t^* r(z) N(z). \tag{5}$$

The tangent modulus $E_t$ for 6th-order-Ogden constitutive model can be derived by differentiating Eq. (2) with respect to $\varepsilon$, and written as

$$E_t(S) = \sum_{i=1}^{6} \frac{2\mu_i}{\alpha_i} \left\{ (\alpha_i - 1)\left[1 + \phi^{-1}(S)\right]^{\alpha_i - 2} + (2\alpha_i + 1)\left[1 + \phi^{-1}(S)\right]^{-2\alpha_i - 2} \right\}, \tag{6}$$

where $\varphi^{-1}(S)$ is the inverse function of $S = \varphi(\varepsilon)$ given by Eq. (2). The mean contact stress $F(z)/A(z)$ is used to evaluate the Piola-Kirchhoff stress $S$ in the current model. Substituting Eq. (6) into Eq. (4), the interfacial stiffness could be denoted as

$$\frac{dF(z)}{dz} = 2r(z)N(z)\sum_{i=1}^{6} \frac{2\mu_i}{\alpha_i} \left\{ (\alpha_i - 1)\left[1 + \phi^{-1}\left(\frac{F(z)}{A(z)}\right)\right]^{\alpha_i - 2} + (2\alpha_i + 1)\left[1 + \phi^{-1}\left(\frac{F(z)}{A(z)}\right)\right]^{-2\alpha_i - 2} \right\}. \tag{7}$$

The geometric functions $A(z)$ and $N(z)$ in Eq. (7) can be obtained either numerically [14] and analytically [42] depending on the specific surface morphology. Subsequently, the relationship between the actual contact area $A$ and the normal load $F$ can be derived by solving the differential equation Eq. (7) using the explicit iteration method with the initial condition $F(z_{\max}) = 0$.

## 5 Results and discussion

Fig. 6 shows the evolution of the actual contact area for Sample T1 obtained through experimental and profile theory methods. Because the acquisition of three-dimensional surface data and rough surface contact experiments were not conducted on the same device, it was necessary to pre-rotate the three-dimensional surface data to align the truncated contact regions with those in the contact experiments. The processing of experimental images employed the Otsu method [43], which is capable of automatically identifying contact and non-contact pixels, and determining the contact area by counting the number of contact pixels. Due to deformation of rough surface, some differences in the obtained contact regions between the two methods are inevitable. However, the consistency in the primary contact regions can be ensured in this work.

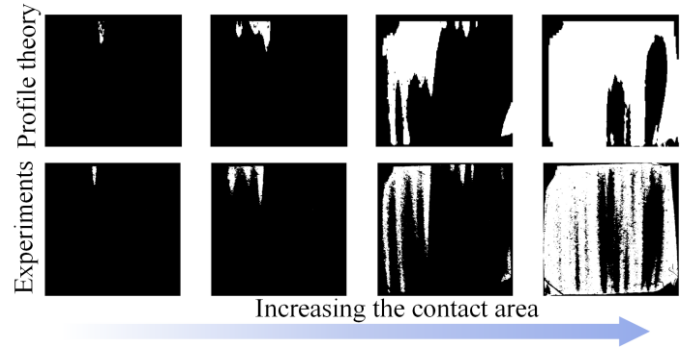

Fig. 6 The evolution of the actual contact area of Sample T1 obtained through experiments and profile theory.

The normalized load-area relationships for samples T1, T2, T3 and T4 are shown in Fig. 7. The dotted lines represent the ICM model calculations and the symbols represent the experimental results. The obvious differences in load-area relationships of four samples could be observed, reflecting that the contact load $F_{T2} > F_{T2} > F_{T3} > F_{T4}$ under the same contact area fraction. This phenomenon implies that the mean contact stresses on the rough surfaces follow the same trend. Additionally, Fig. 7 also reveals that the predictions of the incremental model are in good agreement with the experimental results in a contact fraction range of 90%. However, it must be admitted that there are still some discrepancies between the individual experimental curves and the incremental model results, which may be related to the finite deformation of the rough surface. When the indentation depth is large, the bottom regions of some rough surface asperities no longer maintain the original contour due to finite deformation. The profile theory used in the incremental model assumes that the surface contour of the uncontacted region will remain unchanged throughout the contact process. This assumption may introduce some differences. However, in terms of the overall effect, the incremental model, which takes into account the material nonlinearity of the hyperelastic material, can predict the contact response of the rough surface successfully.

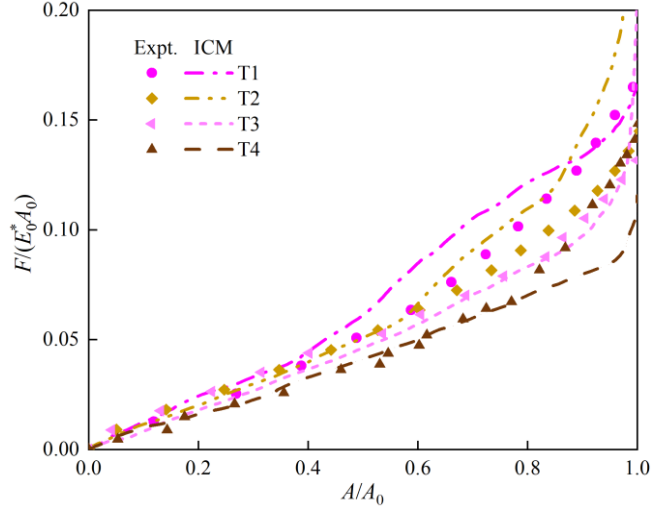

Fig. 7 The evolution of the normalized load $F/(E_0^* A_0)$ with respect to the actual contact area fraction $A/A_0$ for rough surfaces T1, T2, T3, and T4, respectively

In the calculation of load-area relationship, we adopt the instantaneous tangent modulus $E_t$, which depends on the mean contact stress. Then, the dimensionless tangent modulus $E_t^*/E_0^*$ varying with contact area fraction is illustrated in Fig. 8. It is found that the $E_t^*/E_0^*$ falls within the range of 2.2~3, which is similar to the results of the previous research [22] on validating the incremental model using FEM. The $E_t$ in initial stage of contact is larger than in middle stage, it maybe due to the higher contact stress introduced by high frequency of rough surface, which is dominated by material properties. Furthermore, this observation underscores the profound influence of the hyperelastic material nonlinearity in response to contact stress. This influence would increase the instantaneous modulus of the material and the stiffness of contact interface. In contrast, the modulus of conventional linear elastic materials remains unaffected by variations in stress. By introducing the tangent modulus into the incremental model, we effectively incorporate the nonlinearity of hyperelastic materials into numerical calculations.

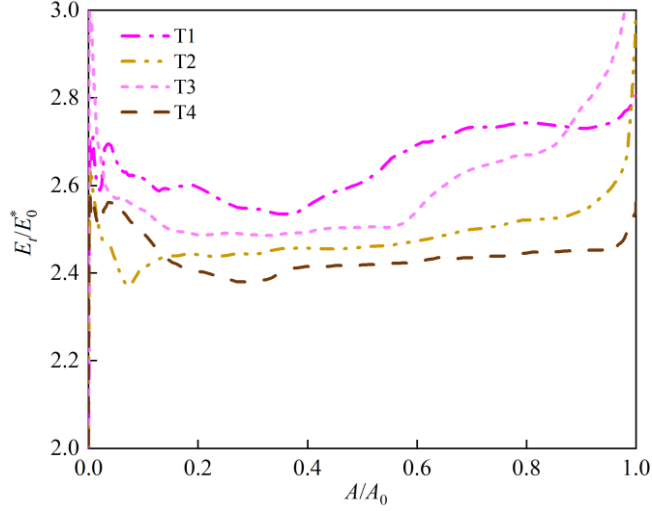

Fig. 8 The normalized composite tangent modulus $E_t^*/E_0^*$ as a function of contact area fraction $A/A_0$ for rough surfaces T1, T2, T3, and T4.

**6 Conclusions**

In this work, the hyperelastic constitutive model is determined by three types of tensile experiments UT, PT, and BT. Four samples were manufactured from the same material batch, the random rough surfaces of which are scanned with a white light interferometer to obtain 3D morphology data. Subsequently, the morphology data were substituted into the incremental model for the hyperelastic material, rendering the relationship between contact load and actual area could be predicted. Finally, we conducted rough surface contact experiments using the optical interferometric technique to validate the modified incremental model.

An agreement between the results of the theoretical model and the experimental method is achieved within a contact fraction range of 90%. Consistent with previous research [22], the findings indicate that the ratio of the tangent modulus to the linear modulus of elasticity, $E_t^*/E_0^*$, will vary with the contact stress and fall within the range of 2.2 ~ 3. This observation underscores that the nonlinear characteristics of hyperelastic materials are amplified due to stress concentration and profoundly influence the contact behaviors of rough surfaces. Moreover, it highlights that the material's nonlinearity can be effectively incorporated into numerical calculations by introducing the tangent modulus into the incremental model. Future research endeavors may aim to enhance and broaden the model's applicability, extending its utility across a broader spectrum of hyperelastic materials and diverse contact scenarios.

**Declaration of competing interest**

The authors have no relevant financial or non-financial interests to disclose.

**Acknowledgment**

This research was funded by the National Natural Science Foundation of China (Grant numbers 12372100).

**Data availability**

All data included in this study are available upon request by contact with the corresponding author.

**Author Contributions**

Chunyun Jiang and Yanbin Zheng contributed to the study's conception and design. Experiments and analysis were performed by Chunyun Jiang. The first draft of the manuscript was written by Chunyun Jiang and all authors commented on previous versions of the manuscript. All authors read and approved the final manuscript.